\documentclass[12pt]{article}
\usepackage[utf8]{inputenc}
\usepackage[top=0.85in,bottom=0.85in,left=1in,right=1in]{geometry}

\usepackage{booktabs} 

\usepackage{authblk} 
\usepackage{graphicx} 
\usepackage{comment}

\usepackage{breakcites}
\usepackage[bookmarks=false]{hyperref}
\usepackage{amsmath}
\usepackage{amssymb}
\usepackage[dvipsnames]{xcolor}
\usepackage{lineno}

\usepackage{tabularx}
\usepackage{array}
\usepackage{longtable}
\newcolumntype{P}[1]{>{\raggedright\arraybackslash}p{#1}}
\newcolumntype{Z}[1]{>{\centering\arraybackslash}p{#1}}

\usepackage{tablefootnote}

\usepackage[style=apa,backend=biber,natbib=true]{biblatex}
\title{Unequal Access to Power in Corruption Networks: Evidence from Colombia}

\author[1]{Giovanna Rodríguez-García}
\author[2]{María Elizabeth Mesa-Pineda} 
\author[3]{José R. Nicolás-Carlock}

\affil[1]{\normalsize Facultad de Economía y Negocios, Universidad Autónoma de Bucaramanga, Bucaramanga, Colombia. E-mail: \url{grodriguez304@unab.edu.co}}
\affil[2]{\normalsize Universidad Nacional de Colombia, Bogota, Colombia. E-mail: \url{mmesap@unal.edu.co}}
\affil[3]{\normalsize Instituto de Física, Universidad Nacional Autónoma de México, Mexico City, Mexico. E-mail: \url{jnicolas@unam.mx}}
\date{}

\begin{document}
\maketitle

\begin{abstract}
Corruption is embedded in networks of access, coordination, and protection, yet little is known about how gender shapes actors’ positions within them. This article examines whether corruption networks in Colombia’s territorial press reproduce gendered patterns of exclusion. Drawing on an access-to-power perspective, we argue that women’s lower presence may reflect unequal incorporation into the spaces where exchanges are organized. Empirically, we use Transparencia por Colombia’s Radiografía de Hechos de Corrupción, integrating case- and actor-level information to build co-participation networks. We analyze gender differences in composition, position, recurrence, and institutional access to resources. Results show that these networks are strongly masculinized: men dominate actors and ties, and women appear less often among recurrent actors. However, women are not absent from connected or dense areas, suggesting uneven rather than absolute exclusion. The findings shift attention from whether women are “less corrupt” to how unequal access to power structures participation in corruption networks.
\end{abstract}


\pagebreak

\section{Introduction}

Corruption is rarely an isolated individual act. It is usually embedded in relational structures that provide access to information and resources, coordination, and protection \citep{Martins2022, waxenecker2025prosecution, lunapla2024network, costa2024nexus, lunapla_nicolas_2020}. Bribes, clientelistic exchanges, illegal contracting schemes, and influence networks depend not only on individual willingness to engage in corrupt behavior, but also on the social and institutional conditions that make such behavior possible \citep{granados_nicolas_2021, kertesz2021complexity, Slingerland2021}. For this reason, corruption can be understood as a networked practice: actors are not only isolated decision-makers but also participants in broader structures of co-participation, brokerage, recurrence, and access to resources.

This relational view is especially relevant for the study of gender and corruption \citep{Pessa2025}. A large body of research has asked whether women are less corrupt than men, often linking women’s lower observed participation in corruption to differences in honesty, risk aversion, prosociality, or tolerance of corruption \citep{Swamy2001,Dollar2001,Chaudhuri2012,Esarey2013,Esarey2017}. Although this literature has generated important findings, it also risks treating gender differences in corruption as if they were mainly the result of individual traits or preferences. A different interpretation emphasizes access to power \citep{Goetz2007,Bjarnegrd2018}. From this perspective, women may appear less frequently in corruption cases not because they are inherently less corrupt, but because they have historically had more limited access to the political offices, bureaucratic positions, informal networks, and resource structures through which corruption is organized and sustained.

This article adopts a structural perspective. If corrupt exchanges depend on trust, institutional discretion, and control over resources, then gendered inequalities in access to power should be reflected within corruption networks. These patterns should not appear only in the number of women and men linked to corruption cases. They should also be visible in how actors are connected, whether they appear across multiple cases, whether they occupy central or cohesive network positions, and whether they hold institutional roles with greater or lesser access to resources.

The article asks: How are women and men differently incorporated into reported corruption networks in Colombia? We analyze three dimensions of gendered access to these networks. First, we examine composition: whether women account for a smaller share of actors and ties than men. Second, we examine position: whether women occupy less central or less cohesive locations in the network, using degree, recurrence, membership in the largest connected component, and $k$-core decomposition. Third, we examine institutional segmentation: whether women, when present, are more likely to occupy actor categories or institutional roles with lower access to discretionary resources. To capture this last dimension, we construct a five-point resource-access weight based on actors’ institutional category and reported position.

The article makes three contributions. First, it contributes to the gender and corruption debate by shifting the focus from whether women are less corrupt to how women and men are differently incorporated into corrupt networks. Second, it contributes to the literature on corruption networks by adding a gendered and institutional dimension to analyses of co-participation, centrality, recurrence, and core structure. Third, it provides evidence from Colombia, a Latin American case where corruption, territorial politics, public contracting, and unequal access to political and administrative power are deeply intertwined.

The article proceeds as follows. Section 2 reviews the classic debate on gender and corruption and develops the access-to-power perspective. Section 3 links this perspective to corruption networks and presents the hypotheses. Section 4 describes the data, network construction, centrality measures, and resource-access coding. Section 5 presents the empirical results. Section 6 discusses the findings in relation to the literature on gender, corruption, and access to power. Section 7 concludes.


\section{The classic debate: Are women less corrupt?}

The first generation of gender and corruption studies found a negative association between the two. \citet{Swamy2001} found that women were less likely to be involved in bribery and less likely to condone bribe-taking, while \citet{Dollar2001} argued that countries with higher female representation in parliament tended to have lower levels of corruption. These studies helped establish the influential “fairer sex” hypothesis, according to which women may be more honest, public-spirited, or less tolerant of corrupt behavior. However, this argument has been strongly challenged. \citet{Sung2003} showed that the association between women’s representation and lower corruption may reflect the broader quality of democratic institutions rather than intrinsic ethical differences between women and men. For this reason, the honesty argument should be treated as an early and contested explanation, rather than as settled evidence.

A second generation of studies shifts attention from ethics to risk preferences. Because corrupt exchanges involve uncertainty, detection, punishment, and reputational costs, women’s greater risk aversion could make them less likely to participate in corruption \citep{Esarey2017}. This argument builds mainly on experimental evidence showing gender differences in risk preferences \citep{Croson2009}.

A closely related behavioral explanation focuses on prosociality. Experimental economics has found that women often exhibit stronger other-regarding preferences, although the evidence is mixed. For example, \citet{Andreoni2001} show that gender differences in altruism are conditional: women are more altruistic when giving is relatively costly, whereas men give more when altruism is cheaper. This literature suggests a possible mechanism through which women may be less willing to engage in corruption when corrupt behavior harms third parties or public goods. \citet{Chaudhuri2012}'s review of experimental evidence makes this link more explicit, arguing that women often behave in more prosocial and less corrupt ways than men, or that no significant gender differences emerge. However, this evidence remains largely centered on individual preferences and behavior, rather than on the broader institutional or social contexts in which corruption occurs.

A further explanation emphasizes gender differences in tolerance of corruption. This hypothesis has been tested using both observational data \citep{Torgler2010} and experimental data \citep{Alatas2009}. While the former shows consistent evidence that women have a greater aversion to corruption, the latter reports mixed results. 

\subsection{Corruption as gendered access to power}

A competing line of research moves away from the claim that women are inherently less corrupt. Rather than treating women as “political cleaners”, this literature suggests that women’s lower presence in corruption may reflect the gendered organization of access to politics and public life \citep{Goetz2007}. From this perspective, corruption is sustained through informal, male-dominated networks that reward homosocial trust and resources, reproduce male power, and systematically exclude women from political access \citep{Bjarnegrd2018}. The access-to-power mechanism, therefore, offers a more structural interpretation: corruption is not only an individual decision, but also a networked practice that requires trust, information, coordination, discretion, and protection.

This argument has been examined across different areas of politics and public life. In elected bodies, corruption and partiality favor clientelism and traditional networks that disadvantage women in political recruitment \citep{Sundstrm2016}. The same logic extends to executive office, where corruption creates informal barriers to women’s recruitment into cabinets \citep{Stockemer2019}. The argument is further refined by \citet{Bauhr2021}, who introduce the concepts of marginalization and network inclusion. Their study of women mayors shows that women executives can reduce corruption risks, especially when newly elected, but that this effect weakens over time. One interpretation is that women initially enter office from outside established corrupt networks, but their later incorporation into existing political and administrative networks may reduce the behavioral or structural differences observed at the beginning.

Recent work adds further nuance to this mechanism by showing that women’s access to power can be both constrained and strategically enabled by corruption. \citet{Armstrong2021} argue that corruption is often sustained by powerful male networks that reinforce women’s exclusion from politics. However, they also show that corruption can sometimes increase women’s access to power when political leaders face accountability pressures and strategically appoint women because they are perceived as cleaner. This suggests that women’s incorporation into power structures may be uneven and politically instrumental: women can be excluded from corrupt networks, but they can also be included when their perceived integrity becomes useful for political survival.

This critique does not imply that gender differences in risk aversion or corruption tolerance are irrelevant. Rather, it suggests that these mechanisms are conditional. For instance, risk aversion should matter especially when corruption is likely to be detected and punished \citep{Esarey2017}. A similar qualification applies to the argument that women are less tolerant of corruption: women may express greater aversion to corruption in contexts where corruption is socially and institutionally stigmatized, especially in democratic settings \citep{Esarey2013}.

This literature suggests that the central question is not whether women are inherently more honest, more risk-averse, or less tolerant of corruption. These mechanisms may operate, but only under specific institutional conditions. The more relevant question is whether women have equal access to the power networks through which corruption becomes possible. Most studies infer gendered exclusion from patterns of representation, appointment, or recruitment. Less clear, however, is whether such exclusion is visible within corruption networks themselves. In other words, if corrupt exchanges depend on informal structures of trust, protection, coordination, and resource control, then gendered access should leave observable traces not only in women’s presence or absence but also in their position within these networks.


\section{Gendered Structure in Corruption Networks}

If women may be less present in corruption because they have historically been excluded from the informal spaces where corrupt exchanges are negotiated, protected, and sustained, then the structure of corruption networks should also reflect this pattern of unequal access. Corrupt exchanges require trust, protection, brokerage, information, discretion, and access to resources, and these resources are unequally distributed \citep{Goetz2007,Sundstrm2016}. If corrupt systems privilege actors already embedded in informal male networks, then women may be excluded from corrupt exchanges and from the powerful offices through which such exchanges are organized. In other words, gender differences in corruption may depend less on fixed individual traits and more on whether women are peripheral to, excluded from, or incorporated into the networks.

Research on corruption networks supports this view by showing that these networks are not random collections of actors, but organized systems of coordination. \citet{Ribeiro2018} show that political corruption in Brazil tends to occur in small groups that are embedded in broader structures where some actors are especially well connected and where separate scandals become linked through shared participants. \citet{Martins2022} and \citet{PRATES2026} extend this evidence to Spain, showing that corruption networks tend to be organized around small circles of cooperation, recurrent participants, and actors who connect otherwise separate schemes. These findings suggest that corruption depends not only on isolated exchanges, but also on brokerage, repeated collaboration, and the capacity to link different actors and cases.

Recent work also shows that gender disparities appear inside corruption networks. \citet{Pessa2025} analyze corruption scandals in Brazil and Spain and find that women represent only 10 percent of actors in the Brazilian network and 20 percent in the Spanish network. These proportions remain stable over time and across scandals of different sizes. Their findings also show that women are not entirely absent from important network positions: once their overall underrepresentation is taken into account, women and men have broadly similar numbers of connections and similar brokerage roles on average. However, men are still overrepresented among the most central actors, and in the Spanish network, ties between men are more frequent than expected. These findings suggest that gendered exclusion may not operate only through women’s absence from corruption networks, but also through unequal access to the most recurrent, central, and protected positions within them.

These findings are important because they provide one of the few direct tests of gendered structure within corruption networks. However, they also leave room for further analysis. Existing evidence tells us whether women are underrepresented and whether they occupy central positions, but less is known about whether their participation is organized through specific institutional roles, sectors, or channels of access. This article extends this literature by examining corruption scandals reported by Colombia’s territorial press and by analyzing not only women’s presence and position in the network, but also their institutional segmentation within it.

Colombia is a relevant case for examining these expectations because corruption is a national and local governance problem \citep{10.1108/JES-03-2021-0148}. At the same time, Colombian politics has historically been shaped by male-dominated access to elected office, public administration, party structures, contracting networks, and territorial brokerage \citep{Buitrago_Aroca_2017, OECD2020GenderEqualityColombia}. This combination makes Colombia a useful case for asking whether corruption networks reproduce gendered patterns of access to power. If women have had more limited access to the informal spaces where resources, protection, and political influence circulate, then these inequalities should be visible not only in who appears in corruption cases, but also in how actors are positioned within reported corruption networks.

Building on previous results reported in the literature, the first expectation concerns the gender composition of corruption networks. If corrupt exchanges depend on male-dominated networks of trust, protection, brokerage, and resource control, women should appear less frequently among the actors involved in reported corruption cases. So that:\\ 

\noindent\textit{\textbf{H1. Gendered composition hypothesis}: Corruption networks reported by Colombia’s territorial press will be masculinized, with women representing a smaller share of actors and ties than men.}\\

This does not mean that women are absent from corruption, nor that men are inherently more corrupt. Rather, it means that the social and institutional spaces where corruption is organized may be more accessible to men.

The second expectation concerns actors’ positions within the network. If women have more limited access to informal spaces of coordination and protection, their exclusion should not be visible only in lower levels of participation, but also in less advantageous network locations. Actors who are more central or located in the network core are more frequently connected to others, more likely to link different cases or groups, and better positioned to participate in recurrent exchanges. Therefore, if corruption networks reproduce gendered access to power, women should occupy less central positions and be less likely to belong to the network core.\\

\noindent\textit{\textbf{H2. Gendered position hypothesis}: Women will occupy less central positions than men and will be less likely to belong to the network core.}\\


The third expectation concerns institutional segmentation within corruption networks. When women appear in these networks, they are expected to occupy positions with lower access to resources. They may enter corruption networks through specific offices, bureaucratic roles, or administrative positions that provide some access to institutional resources, but not necessarily to the protected or central spaces where key corrupt exchanges are coordinated. We therefore hypothesize that: \\

\noindent\textit{\textbf{H3. Gendered segmentation hypothesis}: When women are present in corruption networks, they will be more likely to occupy institutional sectors or roles with lower access to discretionary resources.}


\section{Data and Methods}
To test the previous hypotheses, we built a network based on corruption scandals reported in territorial press data from Colombia. This section describes the data, the network construction, as well as the definitions we use to analyze centrality and influence.

\subsection{Corruption cases data}

Data on corruption cases in Colombia were obtained from Transparencia por Colombia's \emph{Radiografía de Hechos de Corrupción} \citep{transparencia2024radiografia2016_2022}, an open dataset that systematizes corruption cases in Colombia based on territorial press reports and other public sources. The original database includes case-level information, such as the case identifier, title, year, type of corruption, domain of corruption, affected sector, territorial location, and affected institution. It also includes actor-level information, such as the actor identifier, actor name, affiliated institution, type of individual actor, actor subcategory, and reported office or position. Together, these variables allow actors to be systematically linked to specific corruption cases and provide information on their institutional affiliation and position within the recorded corruption events.

Because the hypotheses of this article concern gendered access to power among individual actors, the analysis was restricted to natural persons. Institutional, organizational, and collective actors were therefore excluded from the analytical sample. The database was then cleaned by removing empty entries, observations without a clearly identifiable individual actor, and duplicate records. The original actor-level table contained repeated observations for some actor-case pairs. Because the network construction requires each individual actor to be linked only once to each case, duplicate records were removed using the combination of the case identifier and the actor identifier as the uniqueness criterion. After this preprocessing step, the working dataset contains 4,455 unique actor-case records, corresponding to 1,080 corruption cases and 4,168 unique individual actors over the period 1995-2022. This actor-case structure enables the reconstruction of individual participation across recorded corruption events and the identification of actors who appear in more than one case.

To analyze gender patterns in the reported corruption networks, we created the variable \texttt{sex}, which classifies individual actors into two categories: \texttt{M} and \texttt{F}. This variable was constructed from the names of the individuals and subsequently standardized for the analysis. In this article, these categories are reported as male and female, respectively. This classification should be interpreted as a binary classification inferred from the available names, rather than as a direct measure of self-identified gender. The available information does not allow the identification of non-binary identities or other forms of gender identification.

Two considerations are important for interpreting the data. First, the dataset records actors associated with documented corruption cases; therefore, inclusion in the database does not necessarily imply a final judicial conviction or individual legal responsibility. Second, because the source is constructed from press reports and complementary public sources, the data reflect documented and publicly visible corruption events rather than the full universe of corrupt practices in Colombia. Some actors, cases, or relations may not have been detected or reported, while others may be more visible due to differences in media coverage, institutional attention, the progress of judicial proceedings, or regional reporting capacity. For this reason, the results should be interpreted as patterns of documented co-participation in corruption cases recorded by the source.

\subsection{Network construction}

\begin{figure}[t!]
\includegraphics[width=1.0\textwidth]{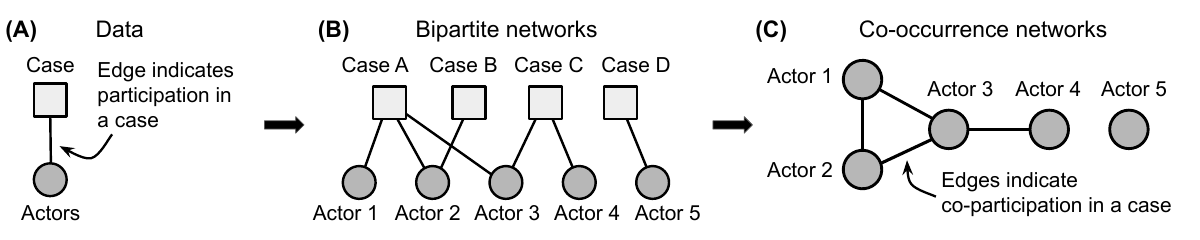}
\caption{\small \textbf{Network construction.} (A) Individual data on cases are treated under an event-actor approach. (B) Aggregated data on cases are modeled as a bipartite network. (C) Co-occurrence networks are created from the co-participation of actors in the same cases.}
\label{fig:fig1}
\end{figure}

We construct the network using a standard case-actor bipartite network approach to corruption cases \citep{Ribeiro2018, Martins2022, Pessa2025}. In this approach, individuals are connected to the cases they appear in, creating a relational structure between two types of nodes: cases and actors (see Fig. \ref{fig:fig1}). We first organize the data into a full case-actor bipartite network, where each tie indicates that an actor was reported to have participated in a given corruption case. We then project this bipartite network onto the set of actor nodes, creating an actor-actor co-participation network. In this projected network, two actors are connected when they appear in the same corruption case. The resulting actor-actor network captures patterns of co-participation among individuals involved in reported corruption events.

Actors that appear in more than one case are of special interest because they connect two or more cases, giving rise to extended networks. Following previous work, we refer to these actors as \textsl{recidivists} \citep{Ribeiro2018, Martins2022, Pessa2025}. Recidivist actors allow the network to extend beyond isolated cases by linking different groups of actors across scandals. For this reason, they are central to understanding whether reported corruption cases remain fragmented or become part of broader relational structures.

\subsection{Network position measures} 

To test the gendered position hypothesis, we examine whether women and men occupy different locations within the actor-actor co-participation network. Previous studies show that actor-actor corruption networks are usually neither fully fragmented nor fully integrated. Instead, they tend to contain one largest connected component (LCC), that is, the largest group of mutually connected actors, alongside several smaller components \citep{Martins2022, Pessa2025}. This structure matters for our analysis because it allows us to distinguish between actors who remain confined to isolated cases and actors who are connected to broader patterns of co-participation across cases.

Our main centrality measure is degree, $k_i$, defined as the number of ties or neighbors of an actor $i$. In our co-occurrence network, degree centrality captures the number of connections an actor (node) has. Higher degree values indicate that an actor has more co-participation ties and is therefore more connected within the reported corruption network. We also report the average degree, defined as the arithmetic mean of degree centrality across actors.

Furthermore, we use a standard core decomposition procedure to examine whether women and men differ in their probability of belonging to more central layers of the network \citep{menczer2020}. Core decomposition uses the degree of each node to separate the network into distinct portions or layers, called \textit{shells}, based on their connectivity levels. Low-degree shells correspond to the periphery, and the densest layer corresponds to the network \textit{core}. More formally, a $k$-core is the maximal subnetwork in which all nodes have degree $k$ or higher within that subnetwork. This measure allows us to evaluate whether women are less likely than men to occupy the most connected and structurally embedded areas of the corruption network.

\subsection{Access to sources}

\newcolumntype{C}[1]{>{\centering\arraybackslash}p{#1}}

\newcolumntype{L}{>{\raggedright\arraybackslash}X}

\newcommand{\headcell}[1]{%
  \begin{tabular}[t]{@{}c@{}}#1\end{tabular}%
}

\begin{table*}[t!]
\centering
\footnotesize
\caption{Actor categories, counts, and gender-weighted percentages.}
\label{tab:actor_categories_gender}

\begin{tabularx}{1.0\textwidth}
{@{} 
L
C{0.08\textwidth}
C{0.08\textwidth}
C{0.08\textwidth}
C{0.16\textwidth}
C{0.16\textwidth}
@{}}
\toprule

\multicolumn{1}{c}{\headcell{\textbf{Actor category}}} &
\headcell{\textbf{Total}\\\textbf{(N)}} &
\headcell{\textbf{Women}\\\textbf{(N)}} &
\headcell{\textbf{Men}\\\textbf{(N)}} &
\headcell{\textbf{Wt. \% Women}\\\textbf{within category}} &
\headcell{\textbf{Wt. \% Men}\\\textbf{within category}} \\

\midrule

Public servant & 1122 & 304 & 818 & 27\% & 63\% \\
Private-sector member & 689 & 142 & 547 & 21\% & 79\% \\
Member of the Public Force & 675 & 12 & 663 & 2\% & 98\% \\
Authority elected by popular vote & 651 & 77 & 574 & 12\% & 88\% \\
Individual actor & 455 & 96 & 359 & 21\% & 79\% \\
Actor linked to an illegal economy or armed group & 293 & 28 & 265 & 10\% & 90\% \\
High-ranking official not elected by popular vote & 140 & 36 & 104 & 26\% & 74\% \\
Member of the third sector & 68 & 26 & 42 & 38\% & 62\% \\
Member of a political organization & 38 & 10 & 28 & 26\% & 74\% \\
Not available or missing & 34 & 9 & 25 & 29\% & 71\% \\
Other & 4 & 1 & 3 & 25\% & 75\% \\
\midrule
Total & 4168 & 741 & 3427 & 18\% & 82\% \\
\bottomrule
\end{tabularx}

\end{table*}

To examine access to sources, we created a set of resource-access categories that measure the extent to which each actor’s institutional location provides access to political, administrative, coercive, or economic resources. These categories were constructed using two variables from the dataset. The first, type of individual actor \textit{(tipo de actor individual)}, identifies the actor’s general location of power by distinguishing broad actor types and subtypes. The second, position \textit{(cargo)}, provides a more specific location within those categories by identifying the office, role, or institutional position held by the actor. Combining these variables produced 78 actor-position categories, which were mapped onto a five-point scale ranging from 1, indicating lower access to resources, to 5, indicating higher access.

For analytical purposes, actor-position codes were grouped into broader actor categories (Table \ref{tab:actor_categories_gender}): public servants, private-sector members, members of the Public Force, authorities elected by popular vote, individual actors, actors linked to an illegal economy or armed group, high-ranking officials not elected by popular vote, members of the third sector, members of political organizations, other actors, and cases with unavailable or missing information. Each family includes actors with different levels of resource access. The classification does not assume that all public, private, or criminal actors occupy equivalent positions of power. For example, a public servant who is the legal representative receives 5 points, while a public servant classified as administrative staff receives 2 points. The complete list is in the Appendix Table \ref{tab:resource_access}.

\section{Results}



\begin{figure}[t!]
\includegraphics[width=1.0\textwidth]{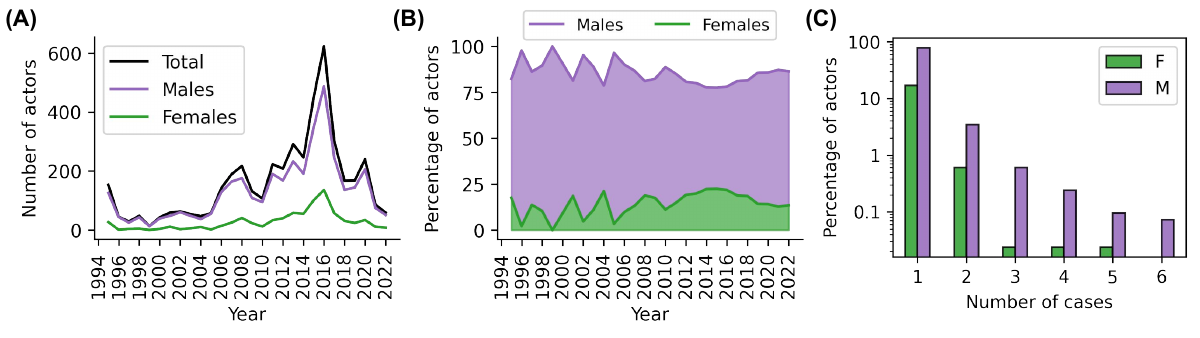}
\caption{\small \textbf{Colombia corruption cases (1995-2022).} (A) Number of actors per year. Among the total of 4168 actors, 3427 (82\%) are males while 741 (18\%) are female. (B) Percentage of actors per year. (C) Distribution of number of cases differentiated by sex. Recidivists (actors that participate in more than one case) constitute only 213 (5\%) of the total.}
\label{fig:fig2}
\end{figure}

The curated dataset of corruption cases in Colombia contains actor-level information on 1,080 cases involving 4,168 actors from the public, private, and criminal sectors between 1995 and 2022 (see Fig. \ref{fig:fig2}A). Of these actors, 3,427 (82\%) are coded as male and 741 (18\%) as female, a gender imbalance that remains relatively stable over time (see Fig. \ref{fig:fig2}B). Recidivists, defined as actors who appear in more than one case, account for 213 individuals, or 5\% of the total. Among them, 185 (87\%) are male, and 28 (13\%) are female (see Fig. \ref{fig:fig2}C). Notably, among the 28 women classified as recidivists, 25 appear in exactly two cases, while the remaining three appear in three, four, and five cases, respectively. These general patterns are consistent with previous studies of corruption networks in Brazil and Spain, where women also represent a minority of actors: approximately 10\% in Brazil and 20\% in Spain \citep{Pessa2025}.

\begin{figure}[t!]
\includegraphics[width=1.0\textwidth]{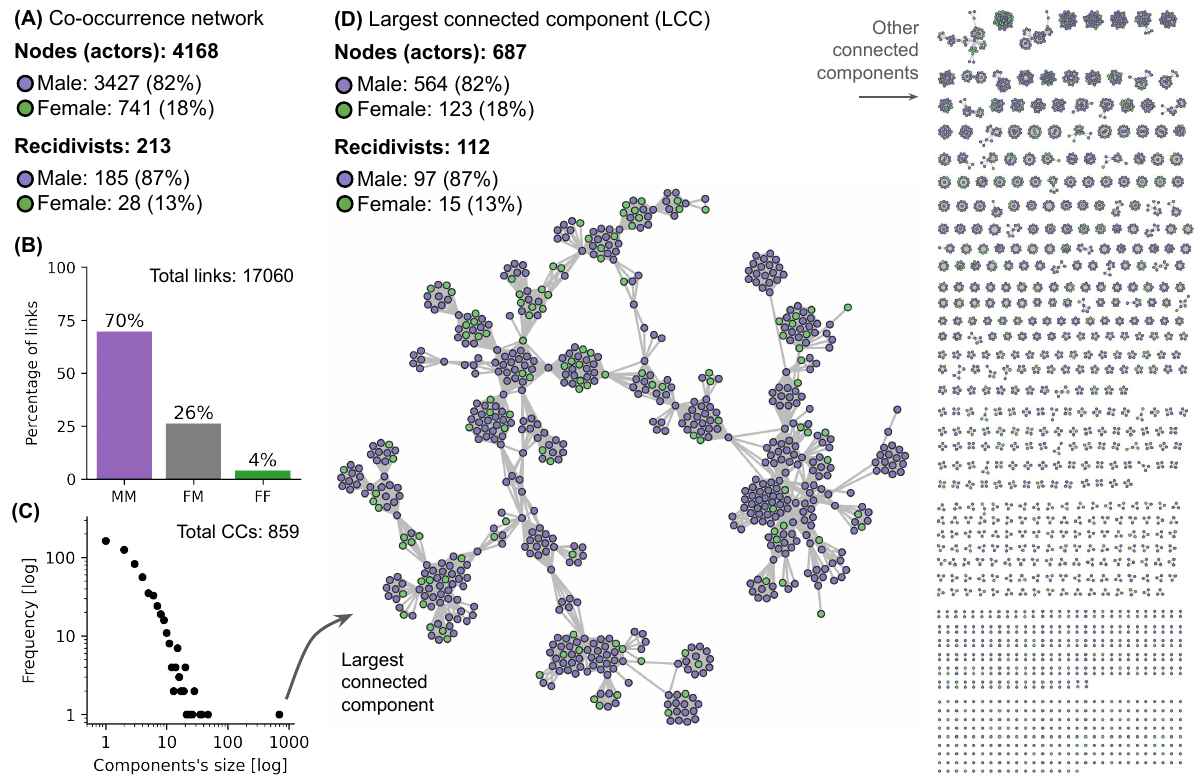}
\caption{\small \textbf{Co-occurrence network.} (A) Node statistics differentiated by sex for all actors and recidivists only. (B) Link distribution differentiated by sex. (C) Connected components' size distribution. (D) Statistics and network visualization of the largest connected component (LCC) along with the rest of the connected components.}
\label{fig:fig3}
\end{figure}

The actor-actor co-participation network exhibits a similar gender imbalance in its tie distribution. Consistent with the actor-level statistics, 70\% of ties are male-male, 26\% are male-female, and only 4\% are female-female (see Fig. \ref{fig:fig3}A-\ref{fig:fig3}B). The network also displays the fragmented structure observed in previous studies of corruption networks. This dataset comprises 859 connected components, or separate subnetworks of mutually connected actors. The largest connected component (LCC) contains 687 nodes, corresponding to 16\% of all actors (see Fig. \ref{fig:fig3}C). Its gender composition closely mirrors that of the full network, with 564 men (82\%) and 123 women (18\%). It also includes 112 recidivists, of whom 97 (82\%) are men, and 15 (18\%) are women (see Fig. \ref{fig:fig3}D). Thus, women’s underrepresentation is not limited to isolated cases; it is also reproduced within the largest connected component of the network.


\begin{figure}[t!]
\includegraphics[width=1.0\textwidth]{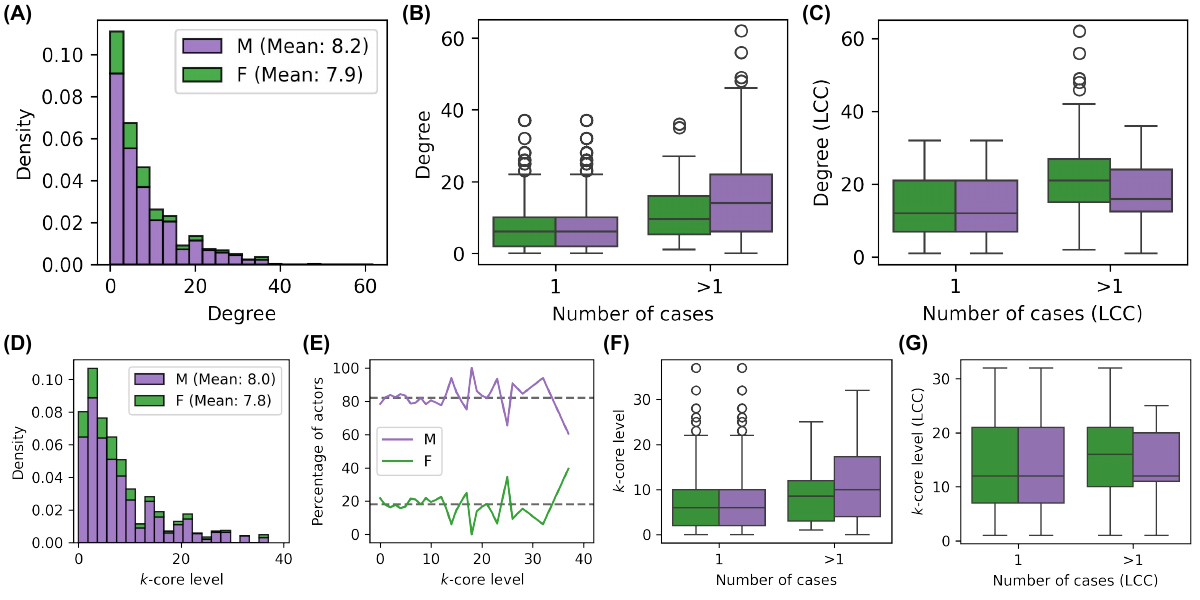}
\caption{\small \textbf{Network position measures.} Measures differentiated by sex. (A) Degree distribution; degree distribution by number of cases for (B) the full network and (C) the LCC. (D) $k$-core level distribution and (E) prevalence of actors according to $k$-core level; $k$-core level distribution by number of cases for (F) the full network and (G) the LCC.}
\label{fig:fig4}
\end{figure}

To evaluate Hypothesis 2, we first examine whether men and women differ in their level of connectivity within the corruption network. Most actors have only a few connections, while a smaller group is connected to many others (see Fig. \ref{fig:fig4}A). When comparing men and women, their average number of connections is very similar, at approximately $k \approx 8$. This suggests that women’s lower presence in the network does not necessarily imply fewer connections among those who do appear. However, differences emerge when we distinguish actors who appear in only one case from those who appear in multiple cases. In the full network, recidivists tend to have more connections than non-recidivists, especially among men (see Fig. \ref{fig:fig4}B). In the LCC, however, the pattern changes: women who appear in more than one case have a slightly higher average number of connections than male recidivists. This indicates that, once women enter the largest connected component, they are not necessarily confined to marginal positions; some are connected to relatively dense areas of co-participation (see Fig. \ref{fig:fig4}C).

The core decomposition points in the same direction. Most actors are located in the lower layers of the network, while only a small number reach the most connected layers (see Fig. \ref{fig:fig4}D). Across most core levels, the shares of men and women remain close to their distributions in the full network. The main exception is the innermost core ($k=37$), where women represent 40\% of actors and men 60\% (see Fig. \ref{fig:fig4}E). This pattern does not support a simple interpretation in which women disappear as network positions become more embedded. Rather, women remain underrepresented in the network overall, but those who enter the largest connected component can also occupy relatively connected and embedded positions. The results by number of cases reinforce this interpretation: in both the full network and the LCC, women who appear in more than one case tend to occupy slightly denser areas of the network than men in the same group (see Fig. \ref{fig:fig4}F-\ref{fig:fig4}G).


\begin{figure}[t!]
\includegraphics[width=1.0\textwidth]{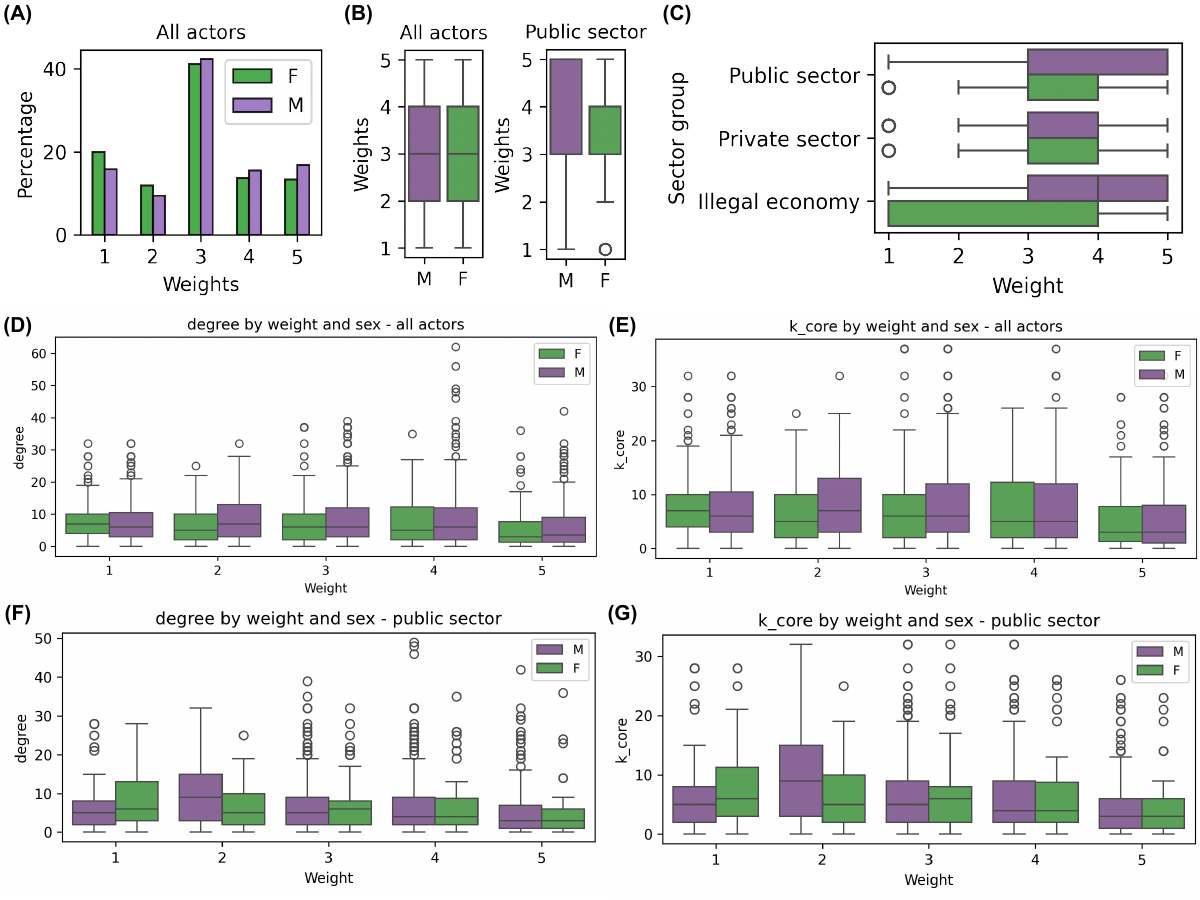}
\caption{\small \textbf{Gendered segmentation.} (A) Weights distribution according to sex. (B) Weights distribution for all and political - public sector actors. (C) Weights distribution for the public sector, private sector and illegal economy. Degree distribution by weight for (D) all actors and (F) public sector actors. $k$-core distribution by weight for (E) all actors and (G) public sector actors.}
\label{fig:fig5}
\end{figure}

To evaluate the gendered segmentation hypothesis (H3), we analyze the distribution of men and women across the resource-access scale. The scale ranges from 1 to 5, with higher values indicating greater potential access to political, administrative, economic, or coercive resources. The results show a moderate but consistent gender difference (see Fig. \ref{fig:fig5}A). Women are more likely to appear in lower-access positions: 31.44\% of women fall into weights 1 and 2, compared with 25.13\% of men. Men, by contrast, are more likely to appear in higher-access positions: 32.10\% of men fall into weights 4 and 5, compared with 26.73\% of women. Although both groups are mainly concentrated in intermediate-access positions, the male distribution is more tilted toward higher levels of institutional and resource access. The average score follows the same pattern, with men reaching an average weight of 3.06 and women 2.85 (see Fig. \ref{fig:fig5}B). 

These differences suggest that women’s underrepresentation is not limited to their lower representation among network actors. It also reflected in the kinds of institutional positions they occupy within reported corruption cases. This pattern becomes clearer when the analysis focuses on political and public-sector actors, whose positions are more directly connected to control over state resources, political representation, administrative decision-making, or institutional brokerage. Within this subset of 1,951 actors, men again have a higher average resource-access score than women: 3.38 compared with 3.15 (see Fig. \ref{fig:fig5}B). The distribution also differs from that of the full network. Women are especially concentrated in intermediate-access positions: 48.48\% of women are located at weight 3, compared with 34.45\% of men. Men, by contrast, are more frequently located in the highest-access category. At weight 5, men represent 25.59\% of cases, compared with 14.52\% of women. This suggests that even within the political and public-sector subset, women are less likely to hold positions with the greatest access to discretionary resources and institutional authority.

The sector-level analysis shows a similar pattern, although the size of the gender gap varies across sectors (see Fig. \ref{fig:fig5}C). In the public sector, where most classified actors are located, men have an average resource-access score of 3.40, compared with 3.16 for women. The difference is smaller in the private sector, where the average score is 3.52 for men and 3.46 for women. The largest gap appears among actors linked to the illegal economy: men have an average score of 3.66, compared with 3.04 for women. This difference should be interpreted cautiously because the number of women in this category is small. These results suggest that gendered segmentation is most visible in the public sector and the illegal economy, while private-sector actors show more similar levels of resource access by gender.

We also examine whether access to resources is associated with actors’ positions in the co-participation network. Among all actors, men generally have similar or higher levels of connection and embeddedness than women at the same resource-access level. The main exception appears at the lowest level of the scale, where women have higher average degree and core values than men (see Fig. \ref{fig:fig5}D-\ref{fig:fig5}E). A similar pattern appears when the analysis is restricted to public-sector actors (see Fig. \ref{fig:fig5}F-\ref{fig:fig5}G). These results suggest that higher-access positions are more often associated with stronger network positions among men than among women. In other words, men’s access to resource-rich roles is more consistently linked to greater involvement in the structure of reported corruption networks.

The statistical comparisons support the descriptive pattern. Among all actors, men have a higher average resource-access score than women by 0.21 points. This difference is statistically significant in both Welch’s test and a two-sided permutation test ($p<0.001$ in both cases). The result remains the same when actors without an assigned positive weight are excluded. The gap is slightly larger among political and public-sector actors, where men’s average score is 0.23 points higher than women’s, and the difference is again significant in both tests ($p<0.001$). For this subset, the Mann--Whitney U test also shows a significant difference in the distribution of weights by sex ($p=0.0006$), even though the median is 3 for both groups. This means that the gender difference is not driven by a shift in the median, but by the way men and women are distributed across the resource-access scale.

At the sector level, the difference is concentrated mainly in the public sector. In this group, men’s average resource-access score is 0.24 points higher than women’s, and the difference is statistically significant in both Welch’s test and the permutation test ($p<0.001$). By contrast, the private-sector gap is small and not statistically distinguishable from zero. Among actors linked to the illegal economy, the average difference is larger, but this result should be interpreted cautiously because the number of women in this category is small and the finding is less stable across tests.

\section{Discussion} 

The findings qualify the access-to-power perspective by showing that gendered exclusion from corruption networks is not simply a matter of absence. The Colombian case analyzed here shows that women are strongly underrepresented among actors involved in corruption cases reported by the territorial press. However, this underrepresentation takes multiple forms. Women are less present in the network, less frequent among actors who appear across multiple cases, and less represented in positions with the greatest access to political, administrative, economic, or coercive resources. At the same time, they are not entirely absent from the more connected or denser areas of the network. This pattern is difficult to explain through arguments centered only on individual honesty, risk aversion, prosociality, or tolerance of corruption. Instead, it is more consistent with the access-to-power critique of the classic gender-and-corruption debate, which emphasizes unequal access to the institutional, informal, and relational spaces in which corrupt exchanges are organized \citep{Goetz2007,Bjarnegrd2018,Sundstrm2016,Bauhr2021}. The discussion below develops this interpretation by revisiting the evidence for each hypothesis.


The first hypothesis expected corruption networks to be masculinized, with women representing a smaller share of actors and ties than men. The results provide strong support for this expectation. Men account for 82\% of all actors, while women account for 18\%. This imbalance remains relatively stable over time and is also reproduced among actors who appear in more than one case (recidivists). The structure of ties follows the same pattern: male--male ties represent 70\% of all ties, male--female ties 26\%, and female--female ties only 4\%. This suggests that the gendered structure of reported corruption networks is not limited to participation in cases, but also extends to patterns of co-participation. In the Colombian cases analyzed here, corruption is reported mainly through male-dominated networks of interaction.

This finding is consistent with previous studies of corruption networks in Brazil and Spain, where women also represent a minority of actors in corruption scandals \citep{Martins2022,Pessa2025}. More importantly, the result supports the access-to-power argument developed in Section 3. Women’s lower presence in corruption networks should not be interpreted as direct evidence that women are inherently less corrupt. Rather, it may reflect their historically unequal incorporation into the informal spaces where political influence, bureaucratic discretion, economic resources, coercive power, and protection circulate. From this perspective, masculinization is not only a compositional outcome. It is also an indicator of unequal access to the networks through which corruption is organized and sustained.


The second hypothesis expected women to occupy less central positions than men and to be less likely to belong to the network core. The evidence provides only partial support for this expectation. The main cluster of linked actors reproduces the gender imbalance observed in the full network: men make up most actors, while women represent a smaller share. This indicates that women are underrepresented in the network, but not disproportionately absent from its largest connected area once their lower presence in the full dataset is taken into account.

The results on network position also complicate a simple interpretation of gendered peripheralization. Men and women have similar average numbers of connections, suggesting that women who appear in reported corruption cases are not necessarily less connected than men. A similar pattern appears across the network’s connected layers: most actors are concentrated in less-connected areas, and the gender composition of these layers largely mirrors that of the full network. Women are also present in the most connected layer \textit{(LLC)}. In addition, among actors who appear in more than one case (recidivists), women in the main cluster show slightly higher average connectivity and are located in somewhat denser areas than men in the same group. This finding should be interpreted cautiously because the number of women in this subgroup is small. Still, it shows that women are not necessarily confined to marginal positions once they enter the more connected parts of the network.

For this reason, H2 receives only partial support. Women are clearly less present in the network and less frequent among actors who appear across multiple cases. However, when they appear in the main cluster or among recurrent participants, they are not always located at the margins. The evidence points to an uneven form of gendered exclusion: the network restricts women’s general access, but it does not fully prevent some women from occupying connected or dense positions. This shifts the interpretation away from a simple exclusion model and toward a pattern of selective incorporation.

This finding connects directly with the distinction between marginalization and network inclusion in the access-to-power literature. Women may be excluded from many informal corruption networks because these networks are often organized through male-dominated spaces of trust, loyalty, brokerage, and protection \citep{Goetz2007,Bjarnegrd2018,Sundstrm2016}. At the same time, women who enter these networks may do so through specific institutional roles, political opportunities, or strategic forms of incorporation \citep{Bauhr2021,Armstrong2021}. The Colombian evidence therefore suggests that women’s lower presence in corruption networks coexists with the selective incorporation of some women into important network positions.

The third hypothesis expected women, when present in corruption networks, to occupy institutional sectors or roles with lower access to discretionary resources. The results provide clearer support for this expectation. The resource-access scale shows that women are more concentrated in lower-access positions and less represented in the highest-access categories. Among women, 31.44\% fall into weight categories 1 and 2, compared with 25.13\% of men. By contrast, 32.10\% of men fall into weight categories 4 and 5, compared with 26.73\% of women. The average score follows the same pattern: 3.06 for men and 2.85 for women.

This difference becomes more visible when the analysis is restricted to political and public-sector actors, whose institutional positions are more directly connected to control over state resources, political representation, administrative decision-making, and institutional brokerage. In this subset, men have an average resource-access score of 3.38, compared with 3.15 for women. Women are especially concentrated in intermediate-access positions: 48.48\% are located at weight 3, compared with 34.45\% of men. Men, by contrast, are more strongly represented at the highest level of the scale, with 25.59\% located at weight 5, compared with 14.52\% of women. This suggests that gendered segmentation is not only present in the full network, but also more pronounced in institutional spaces where control over public resources and decision-making authority is more direct.

The sector-level analysis adds an important qualification. The gender gap is most evident in the public sector, where men have a higher average resource-access score than women. In the private sector, the difference is much smaller, suggesting that gendered access varies across institutional domains. Among actors linked to the illegal economy, the gap is wider, but the small number of women in this category calls for caution. These patterns support the gendered segmentation hypothesis by showing that women’s participation in corruption networks is shaped not only by how often they appear, but also by the institutional positions through which they enter.

The results suggest that gendered access to power leaves three main traces in reported corruption networks. First, it is reflected in composition: men dominate both the population of actors and the structure of ties. Second, it appears in recurrence: men are more likely to appear in multiple cases and, therefore, to connect different corruption events. Third, it is visible in institutional segmentation: women who enter the network tend to occupy positions with lower relative access to resources, especially in political and public-sector spaces. These patterns support the view that corruption is not only an individual decision, but also a networked practice embedded in unequal structures of access.

The results also caution against an overly deterministic interpretation of gendered exclusion. Women’s underrepresentation does not mean that women are absent from corruption or that all women occupy marginal network positions. Some women appear among actors who participate in more than one case, are located in the main cluster of linked actors, or belong to denser areas of the network. This suggests that gendered access to corruption networks may depend on institutional role, sector of participation, control over resources, and relational opportunities within specific cases. Gendered exclusion, therefore, should be understood as a structured tendency rather than an absolute barrier.

\section{Conclusion} 

This article examined whether gendered inequalities in access to power are visible within corruption networks reported by Colombia’s territorial press. Using actor-level data on corruption cases, we analyzed three dimensions of gendered incorporation: who appears in these networks, how actors are positioned, and what institutional roles they occupy. The results show that these networks are strongly masculinized, but not exclusively male. Men represent roughly 80\% of actors across the main network structures, while women represent roughly 20\%. Women are also less likely to appear in multiple cases. However, they do not simply disappear from the more connected areas of the network. Some women who appear in more than one case are relatively well connected, especially within the main cluster of linked actors.

At the same time, the resource-access analysis shows that women are less likely to hold institutional positions with the greatest access to political, administrative, economic, or coercive resources. This pattern is especially visible among political and public-sector actors. These findings are consistent with the access-to-power perspective: corruption networks reproduce gendered inequalities not only by limiting women’s participation but also by shaping the conditions under which they are incorporated into the relational and institutional spaces that enable corrupt exchanges.

These findings contribute to the broader debate on gender and corruption by shifting the focus from whether women are less corrupt to how women and men are differently incorporated into corruption networks. The Colombian evidence supports the argument that women’s lower participation in corruption cases may be linked to unequal access to political and administrative power. It also shows that a network approach can reveal patterns that aggregate representation alone cannot capture. Women’s lower presence, lower recurrence, and concentration in positions with more limited access to resources provide empirical support for the access-to-power mechanism. At the same time, their presence in some dense network locations shows that incorporation into corrupt structures is possible but uneven. In this sense, the article complements previous studies of corruption networks by showing that gender differences are not only compositional or positional, but also institutional: they are associated with the roles and resources through which actors enter corruption networks.


The analysis has several limitations. First, the dataset is based on corruption cases reported by the territorial press and complementary public sources. It therefore captures documented and publicly visible corruption, rather than the full universe of corrupt practices in Colombia. Second, inclusion in the dataset should not be interpreted as legal responsibility or final judicial conviction. Third, the sex variable is inferred from available names and is therefore limited to a binary classification. Fourth, the co-participation network captures actors reported in the same case, but it does not necessarily identify direct collaboration, hierarchy, or intentional coordination among all connected actors. Finally, the resource-access weight is an analytical proxy for institutional access and should be interpreted as an approximation rather than a direct measure of power.


Future research could extend this analysis in several directions. First, longitudinal models could examine whether women’s incorporation into corruption networks changes over time, particularly after institutional reforms, shifts in women’s political representation, or major anti-corruption interventions. Second, null models and permutation tests could assess whether the observed gender composition of ties, cores, and recurrent positions differs from what would be expected given the network’s overall composition. Third, qualitative case studies could examine how women enter specific corruption networks and whether their roles are central, instrumental, symbolic, or subordinate. Finally, comparative analyses with other Latin American countries could determine whether the Colombian pattern reflects a broader regional structure of gendered access to corruption networks.


{\small
\bigskip
\noindent\textbf{Data availability.} Data used in this article is available at Transparencia por Colombia's Radiografía de Hechos de Corrupción \citep{transparencia2024radiografia2016_2022}.

\medskip
\noindent\textbf{Conflict of interests.} The research was conducted in the absence of any commercial or financial relationships that could be construed as a potential conflict of interest.

\medskip
\noindent\textbf{Authors contributions.} GRG: Conceptualization, Writing \& Review, Methodology. MEMP: Data Analysis, Writing \& Review, Methodology. JRNC: Conceptualization, Data Analysis, Writing \& Review, Methodology.

\medskip
\noindent\textbf{Acknowledgements.} JRNC acknowledges support from Mexico's Ministry of Science, Humanities, Technology and Innovation (SECIHTI), under the program ``Estancias Posdoctorales por México 2023''.
}

{\small
\printbibliography
}

\pagebreak

\appendix

\setcounter{table}{0}
\renewcommand{\thetable}{A\arabic{table}}

\renewcommand{\theHtable}{A\arabic{table}}

\section{Resource access coding}
\label{app:resource_access}

\begingroup
\scriptsize
\setlength{\tabcolsep}{4pt}
\renewcommand{\arraystretch}{1.15}

\begin{longtable}
{@{}
P{0.22\textwidth}
P{0.25\textwidth}
Z{0.07\textwidth}
P{0.38\textwidth}
@{}}

\caption{Resource access categories, weights, and coding rationale.}
\label{tab:resource_access} \\

\toprule
\textbf{Actor category} &
\textbf{Actor subcategory} &
\textbf{Weight} &
\textbf{Rationale} \\
\midrule
\endfirsthead

\toprule
\textbf{Actor category} &
\textbf{Actor subcategory} &
\textbf{Weight} &
\textbf{Rationale} \\
\midrule
\endhead

\midrule
\multicolumn{4}{r}{\textit{Continued on next page}} \\
\endfoot

\bottomrule
\endlastfoot

Actor linked to the illegal economy 
& Actor linked to drug trafficking networks 
& 5 
& In Colombia, drug trafficking concentrates substantial resources, corruption capacity, territorial control, and linkages with political, armed, and state networks. \\

Actor linked to the illegal economy 
& Actor linked to arms trafficking 
& 5 
& Arms trafficking implies access to strategic networks, financing, and links with criminal and coercive structures. \\

Actor linked to the illegal economy 
& Actor linked to illegal mining 
& 4 
& Illegal mining usually generates high rents and requires territorial control and coordination with authorities or armed actors. \\

Actor linked to the illegal economy 
& Actor linked to extortion networks 
& 4 
& Extortion implies the systematic extraction of rents and coercive capacity over local actors. \\

Actor linked to the illegal economy 
& Member of Bacrim 
& 4 
& Bacrim operate as organized structures with access to resources, coercion, and territorial control. \\

Actor linked to the illegal economy 
& Member of guerrilla group 
& 4 
& Guerrilla groups have had financing, territorial control, and capacity for intermediation in illegal networks. \\

Actor linked to the illegal economy 
& Actor linked to smuggling networks 
& 3 
& Smuggling generates economic flows and logistical networks, although not always with the same level of control as other illegal economies. \\

Actor linked to the illegal economy 
& Common criminal 
& 1 
& This actor usually has low organizational articulation and lower sustained access to resources or complex corruption networks. \\

Actor linked to the illegal economy 
& No subcategory 
& 4 
& Use when it is not known whether the actor is linked to drug trafficking, illegal mining, Bacrim, smuggling, or other illegal economies. \\

Authority elected by popular vote 
& President 
& 5 
& This is the country’s highest executive authority, with control over the policy agenda, appointments, and state coordination. \\

Authority elected by popular vote 
& Governor 
& 5 
& This position manages departmental budgets, contracting, and bureaucracy, with high territorial power. \\

Authority elected by popular vote 
& Mayor 
& 5 
& This position controls local contracting, budget, and administration, with direct access to public resources. \\

Authority elected by popular vote 
& Congressperson 
& 4 
& This position has high capacity for intermediation, political influence, and access to power networks, although it does not directly execute budgets as a mayor or governor does. \\

Authority elected by popular vote
& Departmental assembly member
& 3
& This position has territorial political influence, but less direct access to resources and lower decision-making capacity than executive or national-level offices. \\

Authority elected by popular vote
& Councilor
& 2
& This position has local influence, but usually less direct control over budgets, contracting, and bureaucracy. \\

Authority elected by popular vote
& No subcategory
& 4
& Use when it is not known whether the actor is a president, governor, mayor, congressperson, departmental assembly member, or councilor. \\

High-ranking official not elected by popular vote
& Minister
& 5
& This position has high decision-making capacity, sectoral coordination, appointment power, and influence over public resources. \\

High-ranking official not elected by popular vote
& Superintendent
& 5
& This position exercises oversight, regulation, and sanctioning powers over strategic sectors, with high institutional access. \\

High-ranking official not elected by popular vote
& Director of Department/Agency
& 5
& This position manages bureaucracy, budget, programs, and high-level administrative decisions. \\

High-ranking official not elected by popular vote
& Director of Autonomous Corporations
& 5
& In Colombia, Regional Autonomous Corporations have influence over licensing, environmental regulation, and territorial resources. \\

High-ranking official not elected by popular vote
& Director of a state-owned enterprise
& 5
& This position can concentrate contracting, budgetary resources, operational decisions, and relevant political connections. \\

High-ranking official not elected by popular vote
& Inspector General of Colombia
& 5
& This position has high disciplinary authority, political visibility, and intervention capacity over public officials and institutions. \\

High-ranking official not elected by popular vote
& Comptroller General of the Republic
& 5
& This position oversees fiscal control over public resources and has national investigative and institutional influence. \\

High-ranking official not elected by popular vote
& National Registrar
& 5
& This position controls key electoral and administrative functions in the Colombian political system. \\

High-ranking official not elected by popular vote
& Attorney General of Colombia
& 5
& This position directs criminal investigations and case prioritization, with broad access to information and institutional power. \\

High-ranking official not elected by popular vote
& Magistrate
& 5
& This position has high legal authority and decision-making capacity, especially in the case of high courts. \\

High-ranking official not elected by popular vote
& University director
& 4
& This position manages budgetary resources and appointments, although with less general political capacity than the highest state offices. \\

High-ranking official not elected by popular vote
& Auditor General of the Republic
& 4
& This position performs relevant oversight functions, although with a narrower scope of power than the Comptroller General's Office or the Inspector General's Office. \\

High-ranking official not elected by popular vote
& Tribunal magistrate
& 4
& This position has relevant judicial authority, but usually lower hierarchy and national influence than a high-court magistrate. \\

High-ranking official not elected by popular vote
& Judicial Branch: senior manager
& 4
& A senior judicial manager may coordinate resources, personnel, or institutional processes with relevant access to administrative decisions. \\

High-ranking official not elected by popular vote
& Municipal ombudsperson
& 3
& This position performs oversight and monitoring functions at the local level, but with a smaller institutional and budgetary scale. \\

High-ranking official not elected by popular vote
& Judicial Branch: manager
& 3
& A judicial manager may handle administrative and budgetary functions, but with less jurisdictional or political power. \\

High-ranking official not elected by popular vote
& No subcategory
& 4
& Use when it is not known whether the actor is a minister, superintendent, magistrate, inspector general, or another high-ranking unelected official. \\

Individual actor
& No subcategory
& 1
& If there is no clear institutional affiliation, this category provides little information about structured access to resources or decision-making networks. \\

Member of political organization
& Director of party/movement/group
& 4
& This position may influence endorsements, alliances, campaign resources, and coordination with public offices. \\

Member of political organization
& Political leader
& 4
& This actor may mobilize support and coordinate power networks, even without holding a formal state office. \\

Member of political organization
& Candidate
& 3
& This actor has potential access to networks, financing, and support, but without consolidated institutional control. \\

Member of political organization
& Party activist/member
& 2
& This actor participates in party structures, but usually has little direct access to strategic resources. \\

Member of political organization
& Sympathizer
& 1
& This actor has a weak political affiliation and limited decision-making capacity or control over resources. \\

Member of political organization
& No subcategory
& 3
& Use when it is not known whether the actor is a director, leader, candidate, party activist/member, or sympathizer. \\

Member of the private sector
& Legal representative
& 5
& This position usually has formal signing authority, legal representation, and decision-making capacity over contracts and resources. \\

Member of the private sector
& Manager
& 4
& This position concentrates coordination, administration, negotiation, and decision-making capacity over strategic processes. \\

Member of the private sector
& Senior executive / director
& 4
& This position indicates high-level command and supervision, with relevant access to organizational decisions. \\

Member of the private sector
& Deputy director
& 3
& This position has medium-high supervision and coordination capacity, but below that of a manager or senior executive. \\

Member of the private sector
& Coordinator
& 3
& This position coordinates processes or teams with intermediate supervisory capacity. \\

Member of the private sector
& Professional staff
& 3
& This position may provide important technical, contractual, or informational access, although with lower formal hierarchy. \\

Member of the private sector
& Administrative staff
& 2
& This position may provide access to procedures, payments, documents, or internal support, but with lower decision-making autonomy. \\

Member of the private sector
& Operational staff
& 1
& This position is more closely associated with execution than with control over resources or decision-making. \\

Member of the private sector
& Assistance/support staff
& 1
& This position is peripheral to resource management, supervision, or decision-making. \\

Member of the private sector
& No subcategory
& 3
& Use when the specific position is not known, such as legal representative, manager, professional staff, administrative staff, or another private-sector role. \\

Member of the Public Force
& Senior officer / director
& 4
& A senior position in the Military Forces or Police implies coordination, supervision, and relevant access to institutional information and resources. \\

Member of the Public Force
& Coordinator
& 3
& This position has intermediate supervisory and coordination capacity, but less autonomy and control than a senior officer or director. \\

Member of the Public Force
& Administrative staff
& 2
& This position may provide access to procedures, documentation, or internal processes, but usually involves less decision-making power. \\

Member of the Public Force
& Professional staff
& 2
& This position may involve technical or specialized access, but usually has lower hierarchy and less control over resources than command positions. \\

Member of the Public Force
& Operational staff
& 2
& This position has execution capacity and territorial presence, but less control over strategic decisions or institutional resources. \\

Member of the Public Force
& No subcategory
& 3
& Use when the specific position or rank is not known. \\

Member of the third sector
& Member of a temporary consortium
& 3
& This actor may participate in organizational arrangements with some coordination of resources or contracts, although with less clarity than in the private or state sector. \\

Member of the third sector
& Social leader
& 2
& This actor may have mobilization and community coordination capacity, but usually lacks direct control over strategic resources. \\

Member of the third sector
& Member of a foundation
& 2
& This actor may participate in organizations with access to projects or agreements, although not necessarily with high decision-making power. \\

Member of the third sector
& Member of an NGO
& 2
& This actor may have intermediation or project implementation capacity, but access to strategic resources is variable and generally lower. \\

Member of the third sector
& Organized civil society: member of a community and/or movement
& 2
& This actor has capacity for mobilization or collective coordination, but access to resources depends on the scope of the organization. \\

Member of the third sector
& Organized civil society: citizen oversight actor
& 2
& This actor has a formal role in oversight and advocacy, but no direct control over public resources. \\

Member of the third sector
& Organized civil society: citizen engaged in social accountability
& 1
& This actor may engage in oversight or reporting, but usually lacks structured access to resources or decisions. \\

Member of the third sector
& No subcategory
& 2
& Use when it is not known whether the actor is a social leader, NGO member, foundation member, citizen oversight actor, or another third-sector actor. \\

Public servant
& Legal representative
& 5
& This position has formal signing authority, representation, and decision-making capacity within the public apparatus. \\

Public servant
& Manager
& 4
& This position usually concentrates administration, coordination, and decision-making over resources, processes, or personnel. \\

Public servant
& Senior official / director
& 4
& This position indicates high-level command, supervision, and coordination within the public entity. \\

Public servant
& District, municipal, and/or departmental secretary
& 4
& In Colombia, these positions usually have relevant access to budgets, implementation, and sectoral political coordination. \\

Public servant
& Deputy director
& 3
& This position has medium-high coordination and supervisory capacity, although less than that of a senior official or director. \\

Public servant
& Coordinator
& 3
& This position coordinates teams or processes with some capacity for intermediation and supervision. \\

Public servant
& Professional staff
& 3
& This position may provide important technical, procedural, or informational access, but with lower formal hierarchy. \\

Public servant
& Administrative staff
& 2
& This position may provide access to documents, procedures, or payments, although with lower autonomous decision-making capacity. \\

Public servant
& Operational staff
& 1
& This position is more closely associated with implementation than with control over resources or strategic decisions. \\

Public servant
& Assistance/support staff
& 1
& This position usually occupies a peripheral location with respect to resource management, supervision, or decision-making. \\

Others
& Member of the media
& 3
& This actor may have access to information, visibility, and public influence, although not stable direct control over public resources. \\

Others
& Member of religious congregations
& 2
& This actor may have legitimacy and community influence, but usually lacks direct control over strategic resources or contracting. \\

Others
& No subcategory
& 3
& Use when the specific position within this residual category is not known. \\

\end{longtable}
\end{document}